\documentclass[11pt, a4paper]{article}
\pdfoutput=1
\usepackage[height=8.85in,width=6.35in]{geometry}

\usepackage{graphicx,rotating}     %usepackage without driver option
\usepackage[bookmarksopen,colorlinks=true,linkcolor=steelblue,
citecolor=darkred,urlcolor=darkred,linktoc=all]{hyperref}

\usepackage{xcolor}
\definecolor{darkred}{rgb}{0.7, 0., 0.}
\definecolor{orangered}{rgb}{1,0.27,0.}
\definecolor{steelblue}{rgb}{0.275,0.51, 0.706}
\definecolor{forestgreen}{rgb}{0.13,0.55,0.13}
\definecolor{violet}{cmyk}{.32,.95,.17,.00}

\usepackage{cite}
\usepackage{amsmath,amssymb,tensor,subdepth}
\usepackage{mathtools}
\usepackage{bm}
\usepackage{bbm}
\usepackage{comment}
\usepackage[utf8]{inputenc}
\usepackage{accents}
\usepackage[footnotesize]{caption}
\usepackage{cancel}
\usepackage{physics}
\usepackage{here}
\usepackage{ascmac}
\usepackage{slashed}
\usepackage{stmaryrd}
\usepackage{trimclip}

\usepackage{tikz}
\usepackage{tikz-feynman}
\tikzfeynmanset{compat=1.1.0}
\pgfdeclarelayer{bg}    % declare background layer
\pgfsetlayers{bg,main}  % set the order of the layers (main is the standard layer)
\usetikzlibrary{decorations}

\pgfdeclaredecoration{dashsoliddouble}{initial}{
  \state{initial}[width=\pgfdecoratedinputsegmentlength]{
    \pgfmathsetlengthmacro\lw{.3pt+.5\pgflinewidth}
    \begin{pgfscope}
      \pgfpathmoveto{\pgfpoint{0pt}{\lw}}%
      \pgfpathlineto{\pgfpoint{\pgfdecoratedinputsegmentlength}{\lw}}%
      \pgfmathtruncatemacro\dashnum{%
        round((\pgfdecoratedinputsegmentlength-3pt)/6pt)
      }
      \pgfmathsetmacro\dashscale{%
        \pgfdecoratedinputsegmentlength/(\dashnum*6pt + 3pt)
      }
      \pgfmathsetlengthmacro\dashunit{3pt*\dashscale}
      \pgfsetdash{{\dashunit}{\dashunit}}{0pt}
      \pgfusepath{stroke}
      \pgfsetdash{}{0pt}
      \pgfpathmoveto{\pgfpoint{0pt}{-\lw}}%
      \pgfpathlineto{\pgfpoint{\pgfdecoratedinputsegmentlength}{-\lw}}%     
      \pgfusepath{stroke}
    \end{pgfscope}
  }
}

\usepackage{chngcntr}
\counterwithin{equation}{section}

\usepackage{libertine}
\usepackage[ISO, upint]{libertinust1math}
\usepackage[scr=boondoxo,bb=boondox]{mathalpha}
\usepackage[T1]{fontenc}

\newcommand{\Mpl}{M_{\text{Pl}}}

\begin{document}
\hypersetup{pageanchor=false}
\begin{titlepage}

\begin{center}
\hfill KEK-TH-2815\\
\hfill CTPU-PTC-26-04\\
\hfill IPMU26-0009 \\
\hfill RESCEU-8/26
\vskip 1.in

\renewcommand{\thefootnote}{\fnsymbol{footnote}}

{\Huge \bf
Revisiting unitarity of single scalar field\\
with non-minimal coupling\\
}

\vskip .8in

{\Large Minxi He$^{(a)}$, Muzi Hong$^{(b)}$, Kyohei Mukaida$^{(c)}$, Tomoya Nishiki$^{(c)}$}

\vskip 0.5in

\begin{tabular}{ll}
$^{(a)}$
&\!\!\!\!\!\emph{Particle Theory and Cosmology Group, Center for Theoretical Physics of the Universe, }\\[-.3em]
& \!\!\!\!\!\emph{Institute for Basic Science (IBS),  Daejeon, 34126, Korea}\\
$^{(b)}$
&\!\!\!\!\!\emph{Department of Physics, Graduate School of Science, }\\[-.3em]
& \!\!\!\!\!\emph{The University of Tokyo, Tokyo 113-0033, Japan}\\
$^{(b)}$& \!\!\!\!\!\emph{Research Center for the Early Universe (RESCEU), Graduate School of Science, }\\[-.3em]
& \!\!\!\!\!\emph{The University of Tokyo, Tokyo 113-0033, Japan}\\
$^{(b)}$& \!\!\!\!\!\emph{Kavli Institute for the Physics and Mathematics of the Universe (WPI), UTIAS,}\\[-.3em]
& \!\!\!\!\!\emph{The University of Tokyo, Kashiwa 277-8583, Japan}\\
$^{(c)}$
& \!\!\!\!\!\emph{Theory Center, IPNS, KEK, 1-1 Oho, Tsukuba, Ibaraki 305-0801, Japan}\\
$^{(c)}$
& \!\!\!\!\!\emph{Graduate University for Advanced Studies (Sokendai),}\\[-.3em]
& \!\!\!\!\!\emph{1-1 Oho, Tsukuba, Ibaraki 305-0801, Japan}\\
\end{tabular}
\end{center}

\vskip .5in
\begin{abstract}
\noindent 
We have investigated the unitarity violation scale of a non-minimally coupled scalar field with quartic self-coupling. 
This model is widely studied in the literature but the estimation of the unitarity violation scale has not been consistently discussed, especially in the Jordan frame. 
We have calculated the six-point scattering amplitudes of the scalar particles in both the Jordan frame and the Einstein frame, and explicitly shown the frame-independence of the results. 
Since the extended target space with the conformal mode is trivial in the single-field case, the dominant contribution comes from the potential of the scalar field. 
The results in both frames become trivial in the vanishing self-coupling limit as expected. 
\end{abstract}

\end{titlepage}

\tableofcontents

\section{Introduction}

Cosmological inflation~\cite{Starobinsky:1980te,Sato:1980yn,Guth:1980zm,Mukhanov:1981xt,Linde:1981mu,Albrecht:1982wi}\footnote{See Ref.~\cite{Sato:2015dga} for a review.} is a widely accepted paradigm, which not only solves the horizon and flatness problems but provides a mechanism to generate the primordial density perturbations.
The current observational data from Planck~\cite{Planck:2018jri} and BICEP/Keck~\cite{BICEP:2021xfz} indicate that the inflationary potential must be concave.
This excludes simple possibilities based on monomial potentials such as $ V \sim \lambda \phi^4$, which naturally arise from quantum corrections unless the inflaton is protected by some symmetries.

An attractive class of models that realizes a concave potential introduces a non-minimal coupling $\xi$ between the inflaton $\phi$ and the Ricci scalar $R$ as
\begin{align}
  \mathcal{L}_\xi = \frac{1}{2} \xi \phi^2 R ~.
\end{align}
This term effectively enhances the Planck scale during inflation for $\phi^2 \gtrsim \Mpl^2 / \xi$ as $M_\text{Pl,eff}^2 \sim  \xi \phi^2$, and thereby effectively reduces the effect of the potential with respect to the Planck scale $\lambda \phi^4 / M_\text{Pl,eff}^4 \sim \lambda / \xi^2$, resulting in a concave potential.
A celebrated example is the ``Higgs inflation,'' where the Standard Model (SM) Higgs field is identified as the inflaton, and its prediction is in excellent agreement with the current data as long as the non-minimal coupling is sizable, $\xi / \sqrt{\lambda} \simeq 4 \times 10^4$~\cite{Cervantes-Cota:1995ehs,Bezrukov:2007ep,Barvinsky:2009ii}.

However, such a large non-minimal coupling induces a strong coupling problem, where the perturbative unitarity is violated at a scale much lower than the Planck scale $\sim \Mpl / \xi$.
This low cutoff scale may jeopardize the validity of the model during inflation and reheating.
Indeed, in the case of Higgs inflation, it has been known that the preheating dynamics exceeds this cutoff scale and therefore the model loses predictability~\cite{Ema:2016dny,Sfakianakis:2018lzf,Ema:2021xhq}.

The estimation of the unitarity violation scale has been extensively discussed in the literature~\cite{Burgess:2009ea,Barbon:2009ya,Burgess:2010zq,Hertzberg:2010dc,Barvinsky:2009ii,Bezrukov:2010jz,Bezrukov:2011sz,Ren:2014sya,Ito:2021ssc,Karananas:2022byw,Mikura:2021clt}, partly because of the subtleties originated from seemingly different descriptions in the Jordan frame and the Einstein frame.
Once we move to the Einstein frame, the inflaton no longer has a non-minimal coupling, but its effect is encoded in the non-canonical kinetic term and the potential. 
In the case of Higgs inflation, since the Higgs fields are a complex doublet which contains four real degrees of freedom (DoFs), the non-minimal coupling is encoded as the non-trivial curvature of the target space spanned by the Higgs fields, which dominantly contributes to the unitarity violation scale in the Einstein frame~\cite{Hertzberg:2010dc,He:2023vlj,He:2023fko}, leading to a cutoff scale $\sim \Mpl / \xi$ consistent with the Jordan frame result.

On the other hand, little is known about the unitarity violation scale in the single-field case.
Contrary to the Higgs inflation, the target space of a single scalar field is trivial, and hence the non-minimal coupling cannot be imprinted in the curvature of the target space.
Moreover, one may readily confirm that the effect of the non-minimal coupling can be completely removed by a Weyl transformation and a field redefinition if the inflaton potential is absent or the non-minimal coupling is conformal, as will be shown later. 
This observation implies that the unitarity violation scale should be expressed as a combination of the coupling in the potential $\lambda$ and the deviation from the conformal coupling $\xi =- 1/6$. 
Nevertheless, the estimation of the unitarity violation scale fulfilling these conditions has not been addressed in the literature, which makes it difficult to understand the equivalence of the Jordan frame and the Einstein frame in this model.
The main goal of this paper is to fill this gap and present a self-consistent treatment in both frames, which explicitly shows the frame-independence of the results.

The paper is organized as follows. 
In Sec.~\ref{sec-setup}, we clarify the model of interest and define the problem that we try to solve in this work. 
We will show that the non-trivial contribution to the unitarity violation scale is expected from the potential of the scalar field instead of the target space. 
We then present the results calculated in the Jordan frame and the Einstein frame in Sec.~\ref{sec-JF} and Sec.~\ref{sec-EF}, respectively, which are shown to be consistent. 
Finally, we summarize our results in Sec.~\ref{sec-conclusion}.

\section{Single scalar field with non-minimal coupling}\label{sec-setup}

In this section, we will lay out the basic setup of the model considered in this work and clarify the problem of unitarity violation scale. 

We consider a non-minimally coupled real scalar field $ \phi $ with quartic self-coupling. 
The defining action of the interested model is written as 
\begin{align}\label{eq-JF-action}
    S_{\rm J} &= \int \dd^4x \sqrt{-g_{\rm J}} \left[ \frac{\Mpl^2}{2} \Omega^2(\phi) R_{\rm J} -\frac{1}{2} g_{\rm J}^{\mu\nu} \partial_{\mu} \phi \partial_{\nu} \phi -V(\phi) \right] ~,~ \Omega^2 \equiv 1+ \xi \frac{\phi^2}{\Mpl^2} ~,~ V \equiv \frac{\lambda}{4} \phi^4 ~,
\end{align}
where $ \lambda $ is the self-coupling constant of $ \phi $, and $ \xi $ denotes the non-minimal coupling between $ \phi $ and gravity and $ \xi =-1/6 $ corresponds to the conformal coupling. 
This model is often considered in the context of the Higgs inflation~\cite{Cervantes-Cota:1995ehs,Bezrukov:2007ep,Barvinsky:2009ii} where $ \phi $ is the inflaton. 
To have successful inflation and fit the observational results from Planck~\cite{Planck:2018jri} and BICEP/Keck~\cite{BICEP:2021xfz}, the combination of the non-minimal coupling and self-coupling should satisfy $ \xi/\sqrt{\lambda} \simeq 4 \times 10^4 $.
If a moderate $ \lambda \sim \mathcal{O}(10^{-2})- \mathcal{O}(10^{-3}) $ is considered, $ \xi $ is then much larger than unity $ \xi \sim \mathcal{O}(10^3) \gg 1 $. 
A question is then raised whether this large non-minimal coupling leads to a cutoff scale much lower than $ \Mpl $ such that it may jeopardize the validity of the model during inflation and reheating. 
Our goal in this work is to find out the cutoff scale in this model in a self-consistent way, in particular around the vacuum $ \phi =0 $. 
Before moving to our main results, we first discuss the approaches in previous works and explain why we need a more careful treatment which is given in this work. 

This question can be investigated in either the Jordan frame or the Einstein frame because the Weyl transformation is just a field-redefinition which should not change physics. 
In the Jordan frame, previous studies~\cite{Bezrukov:2010jz} focus on the non-minimal coupling term $ \xi \phi^2 R_{\rm J} $.
From this operator, they derive the leading order interaction between $ \phi $ and the graviton $ h_{\mu \nu} $ as $ \sim \xi \phi^2 \Box h $ where $ h = h^\mu_\mu $. 
It is tempting to tell from this term that the cutoff scale of the model is obtained by requiring the following scattering amplitude is smaller than unity 
\begin{align}
    \mathcal{M}_{\phi\phi\phi\phi} \sim \frac{\xi^2}{\Mpl^2} E^2 ~,
\end{align}
which gives 
\begin{align}\label{eq-old-LambdaJ}
    \Lambda_{\rm J} \sim \Mpl/\xi ~. 
\end{align}
However, the independence of $ \lambda $ seems to indicate that the potential is playing no role in the determination of $ \Lambda_{\rm J} $, which is actually not true and inconsistent with the Einstein frame results as will be discussed later.

Before going to the Einstein frame, we can understand the importance of the potential by setting $ \lambda =0 $ in Eq.~\eqref{eq-JF-action}. 
Consequently, $ \phi $ is now a massless free field but non-minimally coupled with gravity. 
In this case, if we perform a Weyl transformation 
\begin{align}\label{eq-Weyl-transformation}
    g_{{\rm E} \mu\nu} = \Omega^2 g_{{\rm J} \mu\nu} ~,
\end{align}
which now moves the $ \xi $-dependence into the non-canonical kinetic term as 
\begin{align}
    S_{\rm E} &= \int \dd^4x \sqrt{-g_{\rm E}} \left[ \frac{\Mpl^2}{2} R_{\rm E} -\frac{1}{2} g_{\rm E}^{\mu\nu} \frac{\Omega^2 +6\xi^2 \phi^2/\Mpl^2}{\Omega^4} \partial_{\mu} \phi \partial_{\nu} \phi \right] ~, 
\end{align} 
the theory then is equivalent to a minimally coupled, massless free scalar field with canonical kinetic term as we can further perform a field redefinition 
\begin{align}\label{eq-canonicalization}
    \dd \chi \equiv \sqrt{\frac{\Omega^2 +6\xi^2 \phi^2/\Mpl^2}{\Omega^4}} \dd \phi ~,
\end{align}
to canonically normalize the scalar. 
This canonicalization can be realized because $ \phi $ is a real singlet. 
In the case of the SM Higgs where $ \phi $ is replaced by a complex doublet which contains four real DoFs, the non-trivial intrinsic curvature of the target space spanned by the four DoFs forbids the simultaneous canonicalization of all four fields. 
Inevitably, the target space curvature induces interactions among the DoFs and contributes to the unitarity violation~\cite{Hertzberg:2010dc}. 
A frame-independent method has been developed to straightforwardly calculate such contributions from the geometry of the extended target space~\cite{He:2023vlj,He:2023fko}. 
Here, we focus on the single field case where the (extended) target space is trivially flat. 
As explained above, we do not expect any non-trivial cutoff scale other than the Planck scale from the graviton sector because now $ \xi $-dependence is absent in the Lagrangian, which is in conflict with Eq.~\eqref{eq-old-LambdaJ}. 
This conclusion is consistent with the results in Ref.~\cite{Hertzberg:2010dc} where the author showed that the summation of all contributions purely from the gravity and kinetic sector eventually gives only Planck scale cutoff. 

From above discussion, a non-vanishing $ \lambda $ is essential for $ \xi $ to have non-trivial effects. 
However, that is not clear in the approach used in the literature which only focuses on the non-minimal coupling term and has nothing to do with the scalar potential. 
Even if we set $ \lambda =0 $, we still get the same result, which is obviously in contradiction with our understanding as stated previously. 
On the other hand, the dependence on $ \xi $ in Eq.~\eqref{eq-old-LambdaJ} is also not consistent with the understanding of conformal coupling. 
Specifically, with $ \xi =-1/6 $, the equation of motion for $ \phi $ derived from the action~\eqref{eq-JF-action} is invariant under the transformation $ g_{\mu\nu} \to \omega^2 g_{\mu\nu} $ and $ \phi \to \phi/ \omega $, which means that the dynamics of scalar field is not affected by the scaling of spacetime metric in this case. 
Therefore, we expect that the non-trivial cutoff scale should disappear as $ \xi \to -1/6 $, which is not seen in Eq.~\eqref{eq-old-LambdaJ}. 
Hence, we need to improve the treatment to include the contributions from the potential consistently in the Jordan frame calculation, and the resulting cutoff scale should be eliminated in the limit $ \lambda \to 0 $. 
Before going to our approach, let us have a look at the Einstein-frame calculation. 

Restoring $ \lambda $, we again perform the Weyl transformation~\eqref{eq-Weyl-transformation} and the field redefinition~\eqref{eq-canonicalization} which simplify the gravity sector and the kinetic term by moving the contribution from the non-minimal coupling to the scalar potential. 
The resulting Einstein frame action is then 
\begin{align}\label{eq-EF-action}
    S_{\rm E} &= \int \dd^4x \sqrt{-g_{\rm E}} \left[ \frac{\Mpl^2}{2} R_{\rm E} -\frac{1}{2} g_{\rm E}^{\mu\nu} \partial_{\mu} \chi \partial_{\nu} \chi - U (\chi) \right] ~. 
\end{align}
The non-minimal coupling constant $ \xi $ together with the self-coupling $ \lambda $ now only appears in the potential 
\begin{align}\label{eq-EF-potential}
    U(\chi) \equiv \frac{V}{\Omega^4} (\phi(\chi)) ~,
\end{align}
so we can only focus on the interactions in the potential which may lead to non-trivial cutoff scale. 
Expanding the potential~\eqref{eq-EF-potential} around $ \chi=0 $ up to dimension six, we have 
\begin{align}
    U(\chi) = \frac{\lambda}{4} \delta \chi^4 - \frac{\lambda \xi (1+3\xi)}{3 \Mpl^2} \delta \chi^6 + \mathcal{O}( \delta \chi^8) ~,
\end{align}
from which we can see that the cutoff scale from the dimension-six operator is~\cite{Hertzberg:2010dc,Bezrukov:2010jz}
\begin{align}\label{eq-old-LambdaE}
    \Lambda_{\rm E} = \frac{\Mpl}{4\sqrt{15\lambda \xi (1+3\xi)}} \sim \frac{\Mpl}{\lambda^{1/2} \xi} ~,
\end{align}
where we have used $ \xi \gg 1 $ in the last order estimation. 
We can immediately see the $ \lambda $-dependence in the Einstein frame result~\eqref{eq-old-LambdaE}, contrary to the $ \lambda $-independent $ \Lambda_{\rm J} $ in Eq.~\eqref{eq-old-LambdaJ}. 
In the limit of $ \lambda \to 0 $, $ \Lambda_{\rm E} $ diverges, which is consistent with our previous discussion that $ \lambda $ is essential for $ \xi $ to contribute. 
As argued in Refs.~\cite{Hertzberg:2010dc,Bezrukov:2010jz}, if we take into account higher-order terms such as dimension-$ n $ operators with $ n \gg 4 $, the dependence on $ \lambda $ is going to be suppressed as $ \lambda^{-1/(n-4)} $, which can explain the superficial contradiction. 

From the above discussion, one can see that it is necessary but non-trivial to include the contributions from the potential and recover the $ \lambda $-dependence in the Jordan frame. 
Our goal in this work is to present a self-consistent treatment in both frames and explicitly show the frame independence of the final results.

\section{Jordan frame}\label{sec-JF}

In this section, we are going to show the results calculated from the Jordan frame. 
We will take into account the contributions from both non-minimal coupling and potential. 
The non-minimal coupling between gravity and $ \phi $ is encoded in the scalar part of the spacetime metric, which can be extracted by introducing the conformal mode~\cite{Ema:2020zvg,Ema:2020evi} 
\begin{align}\label{eq-conformal-mode}
    g_{\bullet \mu\nu} = \frac{\Phi_\bullet^2}{6 \Mpl^2} \tilde g_{\mu\nu}, \qquad \det \qty(\tilde g_{\mu\nu}) = -1,
\end{align}
where the black dot implies a subscript associated with the frame, \textit{e.g,} $\bullet = \text{J}, \text{E}$. 
After extracting $ \Phi_{\rm J} $, the action~\eqref{eq-JF-action} becomes 
\begin{equation}\label{eq-JF-action-CM}
    S_{\rm J} =\int \dd^4x\left[\frac{\Phi^2_{\rm J}}{12}\Omega^2 \tilde{R}+\tilde{g}^{\mu\nu}\frac{\Omega^2}{2}\partial_\mu\Phi_{\rm J}\partial_\nu\Phi_{\rm J}-\tilde{g}^{\mu\nu}\frac{\Phi_{\rm J}^2}{12\Mpl^2}\partial_\mu\phi\partial_\nu\phi+\tilde{g}^{\mu\nu}\frac{\xi\phi\Phi_{\rm J}}{\Mpl^2}\partial_\mu\phi\partial_\nu\Phi_{\rm J}-\frac{\Phi_{\rm J}^4}{36\Mpl^4}\frac{\lambda}{4}\phi^4\right] ~. 
\end{equation}
In particular, we are interested in the unitarity violation around some given constant background $ (\Phi_{\rm J}, \phi) = ( \Lambda /\bar{\Omega}, v) $ where $ \bar{\Omega}= \Omega (v) $. 
In this work, we consider the vacuum $ v=0 $, so we expand the action by $ \Phi_{\rm J}= \Lambda + \delta \Phi_{\rm J} $ and $ \phi = 0 + \delta \phi $. 
$ \Lambda $ is then taken to be $ \sqrt{6} \Mpl $ such that the gravitons become strongly coupled at Planck scale in either frame. 
Also, we will focus on $ \phi $ and the trace part of the metric, which is nothing but the conformal mode $ \Phi_\bullet $ and essential to obtain the frame-independent unitarity violation scale below the Planck scale.
As the traceless part is not directly relevant to the frame independence and the lower cutoff scale than the Planck scale, we will not consider the contributions from the traceless part in the following.
As a result, we have perturbed action and we keep terms up to sixth order because our final goal is the tree-level six-point amplitude 
\begin{align}\label{eq-JF-action-perturbation}
    S_{\rm J} \simeq \int \dd^4x \left( \mathcal{L}_{{\rm J},2} + \mathcal{L}_{{\rm J},3} + \mathcal{L}_{{\rm J},4} + \mathcal{L}_{{\rm J},5} + \mathcal{L}_{{\rm J},6} \right) ~,
\end{align}
where 
\begin{align}
    \mathcal{L}_{{\rm J},2} &= \frac{1}{2} \left( \partial \delta \Phi_{\rm J} \right)^2 - \frac{1}{2} \frac{\Lambda^2}{6 \Mpl^2} \left( \partial \delta \phi \right)^2 ~, \nonumber \\
    \mathcal{L}_{{\rm J},3} &= \xi \frac{\Lambda}{\Mpl^2} \delta \phi \,\partial \delta \Phi_{\rm J} \, \partial \delta \phi - \frac{\Lambda}{6\Mpl^2} \delta \Phi_{\rm J} \left( \partial \delta \phi \right)^2 ~, \nonumber \\ 
    \mathcal{L}_{{\rm J},4} &= \frac{\xi}{2\Mpl^2} \delta \phi^2 \left( \partial \delta \Phi_{\rm J} \right)^2 + \frac{\xi}{\Mpl^2} \delta \Phi \, \delta \phi \, \partial \delta \Phi_{\rm J} \, \partial \delta \phi -\frac{1}{12\Mpl^2} \delta \Phi^2 \left( \partial \delta \phi \right)^2 -\frac{\lambda \Lambda^4}{144 \Mpl^4} \delta \phi^4 ~, \nonumber \\ 
    \mathcal{L}_{{\rm J},5} &= -\frac{\lambda \Lambda^3}{36\Mpl^4} \delta \Phi_{\rm J} \delta \phi^4 ~, \nonumber \\ 
    \mathcal{L}_{{\rm J},6} &= -\frac{\lambda \Lambda^2}{24 \Mpl^4} \delta \Phi_{\rm J}^2 \delta \phi^4 ~, \nonumber 
\end{align}
from which we can derive the Feynman rules for the scattering amplitudes used in our calculation. 

Now we will calculate the Feynman diagrams that are relevant for our discussion. 
The contributions to the four-point amplitude $ \delta \phi \delta \phi \to \delta \phi \delta \phi $ are given as follows, respectively, 
\begin{align}
    % contact term
\begin{tikzpicture}[baseline={([yshift=-.5ex]current bounding box.center)}]
  \begin{feynman}
    \vertex (v);
    \vertex [above left=0.6cm of v] (i1);
    \vertex [below left=0.6cm of v] (i2);
    \vertex [above right=0.6cm of v] (f1);
    \vertex [below right=0.6cm of v] (f2);
    \diagram* {
      (i1) -- (v),
      (i2) -- (v),
      (v) -- (f1),
      (v) -- (f2),
    };
  \end{feynman}
\end{tikzpicture}
&= -\frac{i\lambda\Lambda^4}{6\Mpl^4} ~,\label{eq-JF-diagram-4pt-contact} \\
% s-channel
\begin{tikzpicture}[baseline={([yshift=-.5ex]current bounding box.center)}]
  \begin{feynman}
    \vertex (v1);
    \vertex [right=0.8cm of v1] (v2);
    \vertex [above left=0.6cm of v1] (i1) {\footnotesize $1$};
    \vertex [below left=0.6cm of v1] (i2) {\footnotesize $2$};
    \vertex [above right=0.6cm of v2] (f1) {\footnotesize $3$};
    \vertex [below right=0.6cm of v2] (f2) {\footnotesize $4$};
    \diagram* {
      (i1) -- (v1),
      (i2) -- (v1),
      (v1) -- [scalar] (v2),
      (v2) -- (f1),
      (v2) -- (f2),
    };
  \end{feynman}
\end{tikzpicture}
&= -i\left[\frac{1}{p_{12}^2}\frac{\Lambda}{6\Mpl^2}\qty((1+6\xi)p_{12}^2-p_1^2-p_2^2)\frac{\Lambda}{6\Mpl^2}\qty((1+6\xi)p_{12}^2-p_3^2-p_4^2)\right] \label{eq-JF-diagram-4pt-s} ~, 
\end{align}
where the numbers attached to each external line denote the corresponding momentum, and we have adopted the notation $ p_{i \cdots j} = \sum_{n=i}^{j} p_n $. 
For all diagrams in this paper, we assume all momenta attached to the external legs are incoming. 
In our notation, the propagator of $ \delta \phi $ is denoted by a solid line while that for $ \delta \Phi $ is represented by a dashed line, which can be derived from the free part of the Lagrangian $ \mathcal{L}_{{\rm J},2} $ in Eq.~\eqref{eq-JF-action-perturbation} and are shown in Eqs.~\eqref{eq-JF-propagator-Phi} and \eqref{eq-JF-propagator-phi}. 
The coefficients in front of the kinetic terms therefore enters the propagators as the normalization factors. 
One can also define a canonicalized field $ \delta \phi_{\rm J} \equiv \Lambda/(\sqrt{6}\Mpl)\, \delta \phi $ which will be the initial and final states of the scattering processes we consider in the Jordan frame. 
The Feynman rules for the $ n $-point vertices where $ n \geq 3 $ can be derived from the interaction terms in $ \mathcal{L}_{{\rm J},3} $, $ \mathcal{L}_{{\rm J},4} $, $ \mathcal{L}_{{\rm J},5} $, and $ \mathcal{L}_{{\rm J},6} $. 
Since we are finally interested in the tree-level six-point amplitudes with all external legs to be $ \delta \phi $, we will only consider the vertices up to five-point including the conformal mode which are shown in Eqs.~\eqref{eq-JF-3pt} to \eqref{eq-JF-5pt}. 
The diagram in Eq.~\eqref{eq-JF-diagram-4pt-contact} represents the contribution from the four-point contact interaction of $ \delta \phi $ from the potential, which is, therefore, proportional to $ \lambda$, and directly given by the four-point vertex in Eq.~\eqref{eq-JF-4pt-phi4}.  
On the other hand, the diagram in Eq.~\eqref{eq-JF-diagram-4pt-s} is a contribution from the $ s $-channel (if we take $ p_1 $ and $ p_2 $ as the initial momenta) with $ \delta \Phi_{\rm J} $ as an internal line, which is given by the propagator (dashed line) in Eq.~\eqref{eq-JF-propagator-Phi} sandwiched by two three-point vertices in Eq.~\eqref{eq-JF-3pt}. 
As a result, the four-point amplitude is then calculated by the summation of Eq.~\eqref{eq-JF-diagram-4pt-contact} and the $ s $-, $ t $-, and $ u $-channels obtained from Eq.~\eqref{eq-JF-diagram-4pt-s} by switching the momenta as 
\begin{align}\label{eq-JF-4-point}
i\overline{\mathcal{M}}_{\phi\phi\phi\phi}(p_1,p_2,p_3,p_4)
&= 
% contact term
\begin{tikzpicture}[baseline={([yshift=-.5ex]current bounding box.center)}]
  \begin{feynman}
    \vertex (v);
    \vertex [above left=0.6cm of v] (i1);
    \vertex [below left=0.6cm of v] (i2);
    \vertex [above right=0.6cm of v] (f1);
    \vertex [below right=0.6cm of v] (f2);
    \diagram* {
      (i1) -- (v),
      (i2) -- (v),
      (v) -- (f1),
      (v) -- (f2),
    };
  \end{feynman}
\end{tikzpicture}
+
% s-channel
\begin{tikzpicture}[baseline={([yshift=-.5ex]current bounding box.center)}]
  \begin{feynman}
    \vertex (v1);
    \vertex [right=0.8cm of v1] (v2);
    \vertex [above left=0.6cm of v1] (i1) {\footnotesize $1$};
    \vertex [below left=0.6cm of v1] (i2) {\footnotesize $2$};
    \vertex [above right=0.6cm of v2] (f1) {\footnotesize $3$};
    \vertex [below right=0.6cm of v2] (f2) {\footnotesize $4$};
    \diagram* {
      (i1) -- (v1),
      (i2) -- (v1),
      (v1) -- [scalar] (v2),
      (v2) -- (f1),
      (v2) -- (f2),
    };
  \end{feynman}
\end{tikzpicture}
+
% t-channel
\begin{tikzpicture}[baseline={([yshift=-.5ex]current bounding box.center)}]
  \begin{feynman}
    \vertex (v1);
    \vertex [right=0.8cm of v1] (v2);
    \vertex [above left=0.6cm of v1] (i1) {\footnotesize $1$};
    \vertex [below left=0.6cm of v1] (i2) {\footnotesize $3$};
    \vertex [above right=0.6cm of v2] (f1) {\footnotesize $2$};
    \vertex [below right=0.6cm of v2] (f2) {\footnotesize $4$};
    \diagram* {
      (i1) -- (v1),
      (i2) -- (v1),
      (v1) -- [scalar] (v2),
      (v2) -- (f1),
      (v2) -- (f2),
    };
  \end{feynman}
\end{tikzpicture}
+
% u-channel
\begin{tikzpicture}[baseline={([yshift=-.5ex]current bounding box.center)}]
  \begin{feynman}
    \vertex (v1);
    \vertex [right=0.8cm of v1] (v2);
    \vertex [above left=0.6cm of v1] (i1) {\footnotesize $1$};
    \vertex [below left=0.6cm of v1] (i2) {\footnotesize $4$};
    \vertex [above right=0.6cm of v2] (f1) {\footnotesize $3$};
    \vertex [below right=0.6cm of v2] (f2) {\footnotesize $2$};
    \diagram* {
      (i1) -- (v1),
      (i2) -- (v1),
      (v1) -- [scalar] (v2),
      (v2) -- (f1),
      (v2) -- (f2),
    };
  \end{feynman}
\end{tikzpicture}
~,
\end{align}
which, after imposing the on-shell condition, leads to 
\begin{align}
    \overline{\mathcal{M}}_{\phi\phi\phi\phi}=-\frac{\lambda\Lambda^4}{6\Mpl^4} ~.
\end{align}
In Eq.~\eqref{eq-JF-4-point}, the $ s $-, $ t $-, and $ u $-channels are represented by the same diagram with momentum indices exchanged. 
It is clear from this result that the four-point amplitude is controlled by the quartic potential $ \propto \lambda \phi^4 $. 
As long as $ \lambda $ is sufficiently small, there is no unitarity violation. 
To see the contribution from the non-minimal coupling, we have to calculate higher-point amplitudes.
The cancellation among the $ s $-, $ t $-, and $ u $-channels in the four-point amplitude is a consequence of the fact that the target space is flat, which is consistent with our previous discussion and also Ref.~\cite{Hertzberg:2010dc}.

Next, we show the results for the six-point amplitude with six external $ \delta \phi $ legs. 
The relevant Feynman diagrams can be classified into five types as follows: 
\begin{equation}\label{eq-JF-six-point}
\begin{split}
i\overline{\mathcal{M}}_{\phi\phi\phi\phi\phi\phi} =& 
% --- Term 1 (x 15) ---
\sum^{15}
\begin{tikzpicture}[baseline={([yshift=-.5ex]current bounding box.center)}]
  \begin{feynman}
    \vertex (v1) at (0,0);
    \vertex (v2) at (0.8,0);
    \vertex (i1) at (-0.5, 0.4);
    \vertex (i2) at (-0.5, -0.4);
    \vertex (f1) at (1.3, 0.6);
    \vertex (f2) at (1.3, 0.2);
    \vertex (f3) at (1.3, -0.2);
    \vertex (f4) at (1.3, -0.6);
    \diagram* {
      (i1) -- (v1), (i2) -- (v1),
      (v1) -- [scalar] (v2),
      (v2) -- (f1), (v2) -- (f2), (v2) -- (f3), (v2) -- (f4),
    };
  \end{feynman}
\end{tikzpicture}
\;+\;
% --- Term 2 (x 10) ---
\sum^{10}
\begin{tikzpicture}[baseline={([yshift=-.5ex]current bounding box.center)}]
  \begin{feynman}
    \vertex (v1) at (0,0);
    \vertex (v2) at (0.8,0);
    \vertex (i1) at (-0.5, 0.4);
    \vertex (i2) at (-0.6, 0);
    \vertex (i3) at (-0.5, -0.4);
    \vertex (f1) at (1.3, 0.4);
    \vertex (f2) at (1.4, 0);
    \vertex (f3) at (1.3, -0.4);
    \diagram* {
      (i1) -- (v1), (i2) -- (v1), (i3) -- (v1),
      (v1) -- (v2),
      (v2) -- (f1), (v2) -- (f2), (v2) -- (f3),
    };
  \end{feynman}
\end{tikzpicture}
\;+\;
% --- Term 3 (x 60) ---
\sum^{60}
\begin{tikzpicture}[baseline={([yshift=-.5ex]current bounding box.center)}]
  \begin{feynman}
    \vertex (v1) at (0,0);
    \vertex (v2) at (0.8,0);
    \vertex (v3) at (1.6,0);
    \vertex (i1) at (-0.5, 0.4);
    \vertex (i2) at (-0.5, -0.4);
    \vertex (e1) at (1.2, 0.5); % up-left from v2
    \vertex (f1) at (2.1, 0.4);
    \vertex (f2) at (2.2, 0);
    \vertex (f3) at (2.1, -0.4);
    \diagram* {
      (i1) -- (v1), (i2) -- (v1),
      (v1) -- [scalar] (v2),
      (v2) -- (e1),
      (v2) -- (v3),
      (v3) -- (f1), (v3) -- (f2), (v3) -- (f3),
    };
  \end{feynman}
\end{tikzpicture} \\[2em]
% ===== 第二行目 =====
& +\;
% --- Term 4 (x 90) ---
\sum^{90}
\begin{tikzpicture}[baseline={([yshift=-.5ex]current bounding box.center)}]
  \begin{feynman}
    \vertex (v1) at (0,0);
    \vertex (v2) at (0.8,0);
    \vertex (v3) at (1.2,0);
    \vertex (v4) at (2.0,0);
    \vertex (i1) at (-0.5, 0.4);
    \vertex (i2) at (-0.5, -0.4);
    \vertex (e1) at (0.4, 0.5); % up-left from v2
    \vertex (e2) at (1.6, 0.5); % up-right from v3
    \vertex (f1) at (2.5, 0.4);
    \vertex (f2) at (2.5, -0.4);
    \diagram* {
      (i1) -- (v1), (i2) -- (v1),
      (v1) -- [scalar] (v2),
      (v2) -- (e1),
      (v2) -- (v3), % plain (solid)
      (v3) -- (e2),
      (v3) -- [scalar] (v4),
      (v4) -- (f1), (v4) -- (f2),
    };
  \end{feynman}
\end{tikzpicture}
\;+\;
% --- Term 5 (x 45) ---
\sum^{45}
\begin{tikzpicture}[baseline={([yshift=-.5ex]current bounding box.center)}]
  \begin{feynman}
    \vertex (v1) at (0,0);
    \vertex (v2) at (0.8,0);
    \vertex (v3) at (1.6,0);
    \vertex (i1) at (-0.5, 0.4);
    \vertex (i2) at (-0.5, -0.4);
    \vertex (e1) at (0.5, 0.5); % up-left from v2
    \vertex (e2) at (1.1, 0.5); % up-right from v2
    \vertex (f1) at (2.1, 0.4);
    \vertex (f2) at (2.1, -0.4);
    \diagram* {
      (i1) -- (v1), (i2) -- (v1),
      (v1) -- [scalar] (v2),
      (v2) -- (e1), (v2) -- (e2),
      (v2) -- [scalar] (v3),
      (v3) -- (f1), (v3) -- (f2),
    };
  \end{feynman}
\end{tikzpicture} ~,
\end{split}
\end{equation}
where the diagrams include 2-to-4 and 3-to-3 processes with all the initial and final states as $ \delta \phi $. 
Each external $ \delta \phi $ leg is attached with an ingoing momentum $ p_i $ with $ i =1,2, \ldots, 6 $. 
Each type of diagrams can have several different configurations of momenta. 
The sum of contributions over all configurations are symbolically represented by $ \sum^n $ in Eq.~\eqref{eq-JF-six-point} where $ n $ is the number of different configurations. 
The contributions of each type of diagrams with a given momentum configuration shown by numbers attached to each external leg are given as 
\begin{align}
\begin{tikzpicture}[baseline={([yshift=-.5ex]current bounding box.center)}]
  \begin{feynman}
    \vertex (v1) at (0,0);
    \vertex (v2) at (0.8,0);
    \vertex [label=left:{\footnotesize 1}] (i1) at (-0.5, 0.4);
    \vertex [label=left:{\footnotesize 2}] (i2) at (-0.5, -0.4);
    \vertex [label=right:{\footnotesize 3}] (f1) at (1.3, 0.6);
    \vertex [label=right:{\footnotesize 4}] (f2) at (1.3, 0.2);
    \vertex [label=right:{\footnotesize 5}] (f3) at (1.3, -0.2);
    \vertex [label=right:{\footnotesize 6}] (f4) at (1.3, -0.6);
    \diagram* {
      (i1) -- (v1), (i2) -- (v1),
      (v1) -- [scalar] (v2),
      (v2) -- (f1), (v2) -- (f2), (v2) -- (f3), (v2) -- (f4),
    };
  \end{feynman}
\end{tikzpicture}
=i\frac{\lambda\Lambda^4}{9\Mpl^6}(1+6\xi) ~,
\end{align}
\begin{align}
    \begin{tikzpicture}[baseline={([yshift=-.5ex]current bounding box.center)}]
  \begin{feynman}
    \vertex (v1) at (0,0);
    \vertex (v2) at (0.8,0);
    \vertex [label=left:{\footnotesize 1}] (i1) at (-0.5, 0.4);
    \vertex [label=left:{\footnotesize 2}] (i2) at (-0.6, 0);
    \vertex [label=left:{\footnotesize 3}] (i3) at (-0.5, -0.4);
    \vertex [label=right:{\footnotesize 4}] (f1) at (1.3, 0.4);
    \vertex [label=right:{\footnotesize 5}] (f2) at (1.4, 0);
    \vertex [label=right:{\footnotesize 6}] (f3) at (1.3, -0.4);
    \diagram* {
      (i1) -- (v1), (i2) -- (v1), (i3) -- (v1),
      (v1) -- (v2),
      (v2) -- (f1), (v2) -- (f2), (v2) -- (f3),
    };
  \end{feynman}
\end{tikzpicture}
=i\frac{\lambda^2\Lambda^6}{6\Mpl^6}\frac{1}{p_{123}^2} ~,
\end{align}
\begin{align}
    \begin{tikzpicture}[baseline={([yshift=-.5ex]current bounding box.center)}]
  \begin{feynman}
    \vertex (v1) at (0,0);
    \vertex (v2) at (0.8,0);
    \vertex (v3) at (1.6,0);
    \vertex [label=left:{\footnotesize 1}] (i1) at (-0.5, 0.4);
    \vertex [label=left:{\footnotesize 2}] (i2) at (-0.5, -0.4);
    \vertex [label=right:{\footnotesize 3}] (e1) at (1.2, 0.5); % up-left from v2
    \vertex [label=right:{\footnotesize 4}] (f1) at (2.1, 0.4);
    \vertex [label=right:{\footnotesize 5}] (f2) at (2.2, 0);
    \vertex [label=right:{\footnotesize 6}] (f3) at (2.1, -0.4);
    \diagram* {
      (i1) -- (v1), (i2) -- (v1),
      (v1) -- [scalar] (v2),
      (v2) -- (e1),
      (v2) -- (v3),
      (v3) -- (f1), (v3) -- (f2), (v3) -- (f3),
    };
  \end{feynman}
\end{tikzpicture}
=-i\frac{\lambda\Lambda^4}{36\Mpl^6}(1+6\xi)\left[1-(1+6\xi)\frac{p_{12}^2}{p_{456}^2} \right] ~,
\end{align}
\begin{align}
    \begin{tikzpicture}[baseline={([yshift=-.5ex]current bounding box.center)}]
  \begin{feynman}
    \vertex (v1) at (0,0);
    \vertex (v2) at (0.8,0);
    \vertex (v3) at (1.2,0);
    \vertex (v4) at (2.0,0);
    \vertex [label=left:{\footnotesize 1}] (i1) at (-0.5, 0.4);
    \vertex [label=left:{\footnotesize 2}] (i2) at (-0.5, -0.4);
    \vertex [label=right:{\footnotesize 3}] (e1) at (0.4, 0.5); % up-left from v2
    \vertex [label=right:{\footnotesize 4}] (e2) at (1.6, 0.5); % up-right from v3
    \vertex [label=right:{\footnotesize 5}] (f1) at (2.5, 0.4);
    \vertex [label=right:{\footnotesize 6}] (f2) at (2.5, -0.4);
    \diagram* {
      (i1) -- (v1), (i2) -- (v1),
      (v1) -- [scalar] (v2),
      (v2) -- (e1),
      (v2) -- (v3), % plain (solid)
      (v3) -- (e2),
      (v3) -- [scalar] (v4),
      (v4) -- (f1), (v4) -- (f2),
    };
  \end{feynman}
\end{tikzpicture}
=i\frac{\Lambda^2}{216\Mpl^6}(1+6\xi)^2\left[ p_{123}^2-(1+6\xi)(p_{12}^2+p_{56}^2)+(1+6\xi)^2\frac{p_{12}^2p_{56}^2}{p_{123}^2} \right]~,
\end{align}
\begin{align}
    \begin{tikzpicture}[baseline={([yshift=-.5ex]current bounding box.center)}]
  \begin{feynman}
    \vertex (v1) at (0,0);
    \vertex (v2) at (0.8,0);
    \vertex (v3) at (1.6,0);
    \vertex [label=left:{\footnotesize 1}] (i1) at (-0.5, 0.4);
    \vertex [label=left:{\footnotesize 2}] (i2) at (-0.5, -0.4);
    \vertex [label=left:{\footnotesize 3}] (e1) at (0.5, 0.5); % up-left from v2
    \vertex [label=right:{\footnotesize 4}] (e2) at (1.1, 0.5); % up-right from v2
    \vertex [label=right:{\footnotesize 5}] (f1) at (2.1, 0.4);
    \vertex [label=right:{\footnotesize 6}] (f2) at (2.1, -0.4);
    \diagram* {
      (i1) -- (v1), (i2) -- (v1),
      (v1) -- [scalar] (v2),
      (v2) -- (e1), (v2) -- (e2),
      (v2) -- [scalar] (v3),
      (v3) -- (f1), (v3) -- (f2),
    };
  \end{feynman}
\end{tikzpicture}
=i\frac{\Lambda^2}{36\Mpl^6}\xi(1+6\xi)^2(p_{12}^2+p_{56}^2) ~. 
\end{align}
The first contribution above is from a three-point vertex~\eqref{eq-JF-3pt} and a five-point vertex~\eqref{eq-JF-5pt} connected by a $ \delta \Phi_{\rm J} $ propagator. 
The second is the combination of two four-point vertices~\eqref{eq-JF-4pt-phi4}. 
The third comes from a three-point vertex~\eqref{eq-JF-3pt} sandwiched by another three-point vertex and a four-point vertex~\eqref{eq-JF-4pt-phi4}. 
The fourth is a combination of four three-point vertices~\eqref{eq-JF-3pt} such that all six external legs are $ \delta \phi $'s. 
The last diagram is from sandwiching a four-point vertex~\eqref{eq-JF-4pt-phiPhi} by two three-point vertices~\eqref{eq-JF-3pt} with two internal $ \delta \Phi_{\rm J} $ propagators. 

Summing over all momentum configurations as given in Eqs.~\eqref{eq-JF-sum1} to \eqref{eq-JF-sum5} in Appendix~\ref{app-JF}, the final result for six-point on-shell amplitudes for $ \delta \phi $ in the Jordan frame is 
\begin{align}\label{eq-JF-6-point-M}
    \overline{\mathcal{M}}_{\phi\phi\phi\phi\phi\phi}=\frac{\lambda^2\Lambda^6}{6\Mpl^6}\sum^{10}\frac{1}{p_{ijk}^2}+\frac{5\lambda\Lambda^4}{9\Mpl^6} \left(1+6\xi \right)^2 ~,
\end{align}
where we can clearly see $ \lambda $ in both terms, as expected. 
If we take $ \lambda \to 0 $, the result becomes trivial, which is distinct from the results solely calculated from the non-minimal coupling term. 
The combination of $ \left( 1+6\xi \right) $ is also consistent with the understanding of conformal coupling $ \xi =-1/6 $ which minimizes the coupling between gravity and $ \phi $. 
This is one of the main results of this work. 
Now let us see how this result provides the information of unitarity violation. 
To discuss unitarity violation, we focus on the UV regime. 
As $ p_{ijk}^2 \sim s $ where $ s $ is the Mandelstam variable, the first term in Eq.~\eqref{eq-JF-6-point-M} becomes subdominant compared to the second term when we go to high energy limit. 
Focusing on the second term and considering the 2-to-4 processes, we can then obtain the unitarity violation scale by requiring the cross section $ \sigma \lesssim 1/s $ with the cross section calculated as 
\begin{align}
    \sigma \sim \frac{1}{2s} \left| \overline{\mathcal{M}}_{\phi\phi\phi\phi\phi\phi} \right|^2 \Pi_4 \sim s \frac{\lambda^2}{\Mpl^4} \left(1+6\xi \right)^4 ~,
\end{align}
where $ \dd \Pi_4 $ is the four-body phase space of the final states, and we have used $ \Pi_4 \sim s^2 $. 
As a result, we have 
\begin{align}\label{eq-new-LambdaJ}
    \sqrt{s} \lesssim \frac{\Mpl}{ \sqrt{\lambda} \left(1+6\xi \right) } \equiv \Lambda_{\rm J} ~,
\end{align}
whose dependence on $ \lambda $ and $ \xi $ is now consistent with the discussion in Sec.~\ref{sec-setup}. 
Also, the $ \lambda $-dependence is consistent with the results in the literature~\cite{Hertzberg:2010dc,Bezrukov:2010jz}, which is not shown in the conventional Jordan frame calculation. 
Indeed, we expect that $ \lambda $ becomes less important when we consider higher-point amplitudes, but it is essential to point out the necessity of non-vanishing $ \lambda $ in order to have a non-trivial cutoff scale from the non-minimal coupling $ \xi $ in the single-field case. 
In next section, we will compare this with the result in the Einstein frame and show the consistency explicitly, which confirms that our calculation is the self-consistent treatment for this model.

\section{Einstein frame}\label{sec-EF}

Now we go to the Einstein frame by Weyl transformation~\eqref{eq-Weyl-transformation}.
By the definition of the conformal mode~\eqref{eq-conformal-mode}, the Weyl transformation can be achieved by simply rescaling the conformal mode as
\begin{align}
    \Phi^2_{\rm E}=\Omega^2 \Phi^2_{\rm J} ~.
\end{align}
This corresponds to a coordinate transformation of the field basis from $ (\delta \Phi_{\rm J}, \delta \phi_{\rm J}) $ to $ (\delta \Phi_{\rm E},\delta \phi_{\rm E} ) $ where $ \delta \phi_{\rm E} $ is then the canonicalized perturbation in the Einstein frame.  
Around $ (\Phi_{\rm J}, \phi) = ( \Lambda /\bar{\Omega}, v) $, the new field basis in the Einstein frame is then given by a linear transformation 
\begin{align}
    \begin{pmatrix}
        \delta \Phi_{\rm E} \\
        \delta \phi_{\rm E}
    \end{pmatrix}
    =
    \begin{pmatrix}
        ~~\bar{\Omega} & \dfrac{\sqrt{6}\xi v}{\Mpl \bar{\Omega}} \\
        & \\
        ~~0 & \left( 1+ \dfrac{6\xi^2v^2}{\Mpl^2 \bar{\Omega}^2} \right)^{1/2}
    \end{pmatrix}
    \begin{pmatrix}
        \delta \Phi_{\rm J} \\
        \delta \phi_{\rm J}
    \end{pmatrix}
    ~.
\end{align}
When $ v=0 $, this transformation becomes trivial, which explains why the propagators in the Jordan frame (in Appendix~\ref{app-JF}) and the Einstein frame (in Appendix~\ref{app-EF}) are the same. 
This relation is useful to check the consistency of the final results in both frames. 
This will also be useful when constructing a geometric language to achieve a manifestly frame-independent and field-redefinition-independent method in the future where the results should transform covariantly under such coordinate transformation. 

The Einstein frame action is then given by 
\begin{equation}\label{eq-EF-action-CM}
    S=\int\dd^4x\left[\frac{\Phi^2_{\rm E}}{12}\tilde{R}+\frac{1}{2}\tilde{g}^{\mu\nu} \partial_\mu\Phi_{\rm E} \partial_\nu\Phi_{\rm E} -\frac{1}{2}\tilde{g}^{\mu\nu}\qty(\frac{\Phi^2_{\rm E}/\Omega^2}{6\Mpl^2})\qty(1+\frac{6\xi^2}{\Omega^2}\frac{\phi^2}{\Mpl^2})\partial_\mu\phi\partial_\nu\phi-\qty(\frac{\Phi^2_{\rm E}/\Omega^2}{6\Mpl^2})^2\frac{\lambda}{4}\phi^4\right] ~,
\end{equation}
where we can see that the conformal mode is now canonical but the kinetic term and the potential for $ \phi $ contain the information from both $ \xi $ and $ \lambda $. 
Expanding the action by $ \Phi_{\rm E} = \Lambda + \delta \Phi_{\rm E} $ and $ \phi = 0 + \delta \phi $, and keeping terms up to sixth order, we have 
\begin{align}\label{eq-EF-action-perturbation}
    S_{\rm E} \simeq \int \dd^4x \left( \mathcal{L}_{{\rm E},2} + \mathcal{L}_{{\rm E},3} + \mathcal{L}_{{\rm E},4} + \mathcal{L}_{{\rm E},5} + \mathcal{L}_{{\rm E},6} \right) ~,
\end{align}
where 
\begin{align}
    \mathcal{L}_{{\rm E},2} &= \frac{1}{2} \left( \partial \delta \Phi_{\rm E} \right)^2 - \frac{1}{2} \frac{\Lambda^2}{6 \Mpl^2} \left( \partial \delta \phi \right)^2 ~, \nonumber \\
    \mathcal{L}_{{\rm E},3} &= - \frac{\Lambda}{6\Mpl^2} \delta \Phi_{\rm E} \left( \partial \delta \phi \right)^2 ~, \nonumber \\ 
    \mathcal{L}_{{\rm E},4} &= -\frac{1}{12\Mpl^2} \delta \Phi_{\rm E}^2 \left( \partial \delta \phi \right)^2 -\frac{\Lambda^2}{12 \Mpl^4} \xi \left( 6\xi-1 \right) \delta \phi^2 \left( \partial \delta \phi \right)^2 -\frac{\lambda \Lambda^4}{144 \Mpl^4} \delta \phi^4 ~, \nonumber \\ 
    \mathcal{L}_{{\rm E},5} &= -\frac{\Lambda}{6\Mpl^4} \xi \left( 6\xi -1 \right) \delta \Phi_{\rm E} \delta \phi^2 \left( \partial \delta \phi \right)^2 -\frac{\lambda \Lambda^3}{36\Mpl^4} \delta \Phi_{\rm E} \delta \phi^4 ~, \nonumber \\ 
    \mathcal{L}_{{\rm E},6} &= \frac{\xi^2 \Lambda^2}{12 \Mpl^6} \left( 12\xi -1 \right) \delta \phi^4 \left( \partial \delta \phi \right)^2 -\frac{1}{12\Mpl^4} \xi \left( 6\xi -1 \right) \delta \Phi_{\rm E}^2 \delta \phi^2 \left( \partial \delta \phi \right)^2 -\frac{\lambda \Lambda^2}{24 \Mpl^4} \delta \Phi_{\rm E}^2 \delta \phi^4 + \frac{\lambda \xi \Lambda^4}{72\Mpl^6} \delta \phi^6 ~, \nonumber 
\end{align}
from which we derive the Feynman rules in the Einstein frame as shown in Appendix~\ref{app-EF}.

We calculate the six-point amplitude for $ \delta \phi $ to compare with the Jordan frame result in the last section. 
In the Einstein frame, we have six types of diagrams including one contact interaction as follows: 
\begin{equation}\label{eq-EF-six-point}
\begin{split}
i\overline{\mathcal{M}}_{\phi\phi\phi\phi\phi\phi}=& 
\begin{tikzpicture}[baseline={([yshift=-.5ex]current bounding box.center)}]
  \begin{feynman}
    \vertex (v);
    \vertex (l1) at (150:0.6);
    \vertex (l2) at (90:0.6);
    \vertex (l3) at (30:0.6);
    \vertex (l4) at (330:0.6);
    \vertex (l5) at (270:0.6);
    \vertex (l6) at (210:0.6);
    \diagram* {
      (l1) -- [plain](v),
      (l2) -- [plain](v),
      (l3) -- [plain](v),
      (l4) -- [plain](v),
      (l5) -- [plain](v),
      (l6) -- [plain](v),
    };
  \end{feynman}
\end{tikzpicture}
+
% --- Term 1 (x 15) ---
\sum^{15}
\begin{tikzpicture}[baseline={([yshift=-.5ex]current bounding box.center)}]
  \begin{feynman}
    \vertex (v1) at (0,0);
    \vertex (v2) at (0.8,0);
    \vertex (i1) at (-0.5, 0.4);
    \vertex (i2) at (-0.5, -0.4);
    \vertex (f1) at (1.3, 0.6);
    \vertex (f2) at (1.3, 0.2);
    \vertex (f3) at (1.3, -0.2);
    \vertex (f4) at (1.3, -0.6);
    \diagram* {
      (i1) -- (v1), (i2) -- (v1),
      (v1) -- [scalar] (v2),
      (v2) -- (f1), (v2) -- (f2), (v2) -- (f3), (v2) -- (f4),
    };
  \end{feynman}
\end{tikzpicture}
\;+\;
% --- Term 2 (x 10) ---
\sum^{10}
\begin{tikzpicture}[baseline={([yshift=-.5ex]current bounding box.center)}]
  \begin{feynman}
    \vertex (v1) at (0,0);
    \vertex (v2) at (0.8,0);
    \vertex (i1) at (-0.5, 0.4);
    \vertex (i2) at (-0.6, 0);
    \vertex (i3) at (-0.5, -0.4);
    \vertex (f1) at (1.3, 0.4);
    \vertex (f2) at (1.4, 0);
    \vertex (f3) at (1.3, -0.4);
    \diagram* {
      (i1) -- (v1), (i2) -- (v1), (i3) -- (v1),
      (v1) -- (v2),
      (v2) -- (f1), (v2) -- (f2), (v2) -- (f3),
    };
  \end{feynman}
\end{tikzpicture}
\;+\;
% --- Term 3 (x 60) ---
\sum^{60}
\begin{tikzpicture}[baseline={([yshift=-.5ex]current bounding box.center)}]
  \begin{feynman}
    \vertex (v1) at (0,0);
    \vertex (v2) at (0.8,0);
    \vertex (v3) at (1.6,0);
    \vertex (i1) at (-0.5, 0.4);
    \vertex (i2) at (-0.5, -0.4);
    \vertex (e1) at (1.2, 0.5); % up-left from v2
    \vertex (f1) at (2.1, 0.4);
    \vertex (f2) at (2.2, 0);
    \vertex (f3) at (2.1, -0.4);
    \diagram* {
      (i1) -- (v1), (i2) -- (v1),
      (v1) -- [scalar] (v2),
      (v2) -- (e1),
      (v2) -- (v3),
      (v3) -- (f1), (v3) -- (f2), (v3) -- (f3),
    };
  \end{feynman}
\end{tikzpicture} \\[2em]
& +\;
% --- Term 4 (x 90) ---
\sum^{90}
\begin{tikzpicture}[baseline={([yshift=-.5ex]current bounding box.center)}]
  \begin{feynman}
    \vertex (v1) at (0,0);
    \vertex (v2) at (0.8,0);
    \vertex (v3) at (1.2,0);
    \vertex (v4) at (2.0,0);
    \vertex (i1) at (-0.5, 0.4);
    \vertex (i2) at (-0.5, -0.4);
    \vertex (e1) at (0.4, 0.5); % up-left from v2
    \vertex (e2) at (1.6, 0.5); % up-right from v3
    \vertex (f1) at (2.5, 0.4);
    \vertex (f2) at (2.5, -0.4);
    \diagram* {
      (i1) -- (v1), (i2) -- (v1),
      (v1) -- [scalar] (v2),
      (v2) -- (e1),
      (v2) -- (v3), % plain (solid)
      (v3) -- (e2),
      (v3) -- [scalar] (v4),
      (v4) -- (f1), (v4) -- (f2),
    };
  \end{feynman}
\end{tikzpicture}
\;+\;
% --- Term 5 (x 45) ---
\sum^{45}
\begin{tikzpicture}[baseline={([yshift=-.5ex]current bounding box.center)}]
  \begin{feynman}
    \vertex (v1) at (0,0);
    \vertex (v2) at (0.8,0);
    \vertex (v3) at (1.6,0);
    \vertex (i1) at (-0.5, 0.4);
    \vertex (i2) at (-0.5, -0.4);
    \vertex (e1) at (0.5, 0.5); % up-left from v2
    \vertex (e2) at (1.1, 0.5); % up-right from v2
    \vertex (f1) at (2.1, 0.4);
    \vertex (f2) at (2.1, -0.4);
    \diagram* {
      (i1) -- (v1), (i2) -- (v1),
      (v1) -- [scalar] (v2),
      (v2) -- (e1), (v2) -- (e2),
      (v2) -- [scalar] (v3),
      (v3) -- (f1), (v3) -- (f2),
    };
  \end{feynman}
\end{tikzpicture}
~,
\end{split}
\end{equation}
where all the initial and final states are $ \delta \phi $. 
For each type of diagrams, every external $ \delta \phi $ leg is attached with an ingoing momentum $ p_i $ with $ i =1,2, \ldots, 6 $ and different momentum configurations correspond to different contributions to the amplitude.
Again, the sum of contributions over all configurations are symbolically represented by $ \sum^n $ in Eq.~\eqref{eq-EF-six-point} where $ n $ is the number of all possible different configurations as in the Jordan frame. 
The contributions of each type of diagrams with a given momentum configuration shown by numbers attached to each external leg are given as 
\begin{align}
\begin{tikzpicture}[baseline={([yshift=-.5ex]current bounding box.center)}]
  \begin{feynman}
    \vertex (v);
    \vertex (l1) at (150:0.8) {\footnotesize $1$};
    \vertex (l2) at (90:0.8) {\footnotesize $2$};
    \vertex (l3) at (30:0.8) {\footnotesize $3$};
    \vertex (l4) at (330:0.8) {\footnotesize $4$};
    \vertex (l5) at (270:0.8) {\footnotesize $5$};
    \vertex (l6) at (210:0.8) {\footnotesize $6$};
    \diagram* {
      (l1) -- [plain](v),
      (l2) -- [plain](v),
      (l3) -- [plain](v),
      (l4) -- [plain](v),
      (l5) -- [plain](v),
      (l6) -- [plain](v),
    };
  \end{feynman}
\end{tikzpicture}
&=i\frac{10\xi\lambda\Lambda^4}{\Mpl^6} ~,
\end{align}
\begin{align}
\begin{tikzpicture}[baseline={([yshift=-.5ex]current bounding box.center)}]
  \begin{feynman}
    \vertex (v1) at (0,0);
    \vertex (v2) at (0.8,0);
    \vertex [label=left:{\footnotesize 1}] (i1) at (-0.5, 0.4);
    \vertex [label=left:{\footnotesize 2}] (i2) at (-0.5, -0.4);
    \vertex [label=right:{\footnotesize 3}] (f1) at (1.3, 0.6);
    \vertex [label=right:{\footnotesize 4}] (f2) at (1.3, 0.2);
    \vertex [label=right:{\footnotesize 5}] (f3) at (1.3, -0.2);
    \vertex [label=right:{\footnotesize 6}] (f4) at (1.3, -0.6);
    \diagram* {
      (i1) -- (v1), (i2) -- (v1),
      (v1) -- [scalar] (v2),
      (v2) -- (f1), (v2) -- (f2), (v2) -- (f3), (v2) -- (f4),
    };
  \end{feynman}
\end{tikzpicture}
=i\frac{\xi\Lambda^2}{18\Mpl^6}(1-6\xi)p_{12}^2+i\frac{\lambda\Lambda^4}{9\Mpl^6} ~,
\end{align}
\begin{align}
    \begin{tikzpicture}[baseline={([yshift=-.5ex]current bounding box.center)}]
  \begin{feynman}
    \vertex (v1) at (0,0);
    \vertex (v2) at (0.8,0);
    \vertex [label=left:{\footnotesize 1}] (i1) at (-0.5, 0.4);
    \vertex [label=left:{\footnotesize 2}] (i2) at (-0.6, 0);
    \vertex [label=left:{\footnotesize 3}] (i3) at (-0.5, -0.4);
    \vertex [label=right:{\footnotesize 4}] (f1) at (1.3, 0.4);
    \vertex [label=right:{\footnotesize 5}] (f2) at (1.4, 0);
    \vertex [label=right:{\footnotesize 6}] (f3) at (1.3, -0.4);
    \diagram* {
      (i1) -- (v1), (i2) -- (v1), (i3) -- (v1),
      (v1) -- (v2),
      (v2) -- (f1), (v2) -- (f2), (v2) -- (f3),
    };
  \end{feynman}
\end{tikzpicture}
=i\frac{\xi^2\Lambda^2}{6\Mpl^6}(1-6\xi)^2p_{123}^2-i\frac{\xi\lambda\Lambda^4}{3\Mpl^6}(1-6\xi)+i\frac{\lambda^2\Lambda^6}{6\Mpl^6}\frac{1}{p_{123}^2} ~,
\end{align}
\begin{align}
    \begin{tikzpicture}[baseline={([yshift=-.5ex]current bounding box.center)}]
  \begin{feynman}
    \vertex (v1) at (0,0);
    \vertex (v2) at (0.8,0);
    \vertex (v3) at (1.6,0);
    \vertex [label=left:{\footnotesize 1}] (i1) at (-0.5, 0.4);
    \vertex [label=left:{\footnotesize 2}] (i2) at (-0.5, -0.4);
    \vertex [label=right:{\footnotesize 3}] (e1) at (1.2, 0.5); % up-left from v2
    \vertex [label=right:{\footnotesize 4}] (f1) at (2.1, 0.4);
    \vertex [label=right:{\footnotesize 5}] (f2) at (2.2, 0);
    \vertex [label=right:{\footnotesize 6}] (f3) at (2.1, -0.4);
    \diagram* {
      (i1) -- (v1), (i2) -- (v1),
      (v1) -- [scalar] (v2),
      (v2) -- (e1),
      (v2) -- (v3),
      (v3) -- (f1), (v3) -- (f2), (v3) -- (f3),
    };
  \end{feynman}
\end{tikzpicture}
=-i\frac{\Lambda^2}{36\Mpl^6}\qty(\lambda\Lambda^2\qty(1-\frac{p_{12}^2}{p_{123}^2})-\xi(1-6\xi)(p_{123}^2-p_{12}^2)) ~,
\end{align}
\begin{align}
    \begin{tikzpicture}[baseline={([yshift=-.5ex]current bounding box.center)}]
  \begin{feynman}
    \vertex (v1) at (0,0);
    \vertex (v2) at (0.8,0);
    \vertex (v3) at (1.2,0);
    \vertex (v4) at (2.0,0);
    \vertex [label=left:{\footnotesize 1}] (i1) at (-0.5, 0.4);
    \vertex [label=left:{\footnotesize 2}] (i2) at (-0.5, -0.4);
    \vertex [label=right:{\footnotesize 3}] (e1) at (0.4, 0.5); % up-left from v2
    \vertex [label=right:{\footnotesize 4}] (e2) at (1.6, 0.5); % up-right from v3
    \vertex [label=right:{\footnotesize 5}] (f1) at (2.5, 0.4);
    \vertex [label=right:{\footnotesize 6}] (f2) at (2.5, -0.4);
    \diagram* {
      (i1) -- (v1), (i2) -- (v1),
      (v1) -- [scalar] (v2),
      (v2) -- (e1),
      (v2) -- (v3), % plain (solid)
      (v3) -- (e2),
      (v3) -- [scalar] (v4),
      (v4) -- (f1), (v4) -- (f2),
    };
  \end{feynman}
\end{tikzpicture}
=i\frac{\Lambda^2}{216\Mpl^6}\qty(p_{123}^2-p_{56}^2-p_{12}^2+\frac{p_{12}^2p_{56}^2}{p_{123}^2}) ~,
\end{align}
\begin{align}
    \begin{tikzpicture}[baseline={([yshift=-.5ex]current bounding box.center)}]
  \begin{feynman}
    \vertex (v1) at (0,0);
    \vertex (v2) at (0.8,0);
    \vertex (v3) at (1.6,0);
    \vertex [label=left:{\footnotesize 1}] (i1) at (-0.5, 0.4);
    \vertex [label=left:{\footnotesize 2}] (i2) at (-0.5, -0.4);
    \vertex [label=left:{\footnotesize 3}] (e1) at (0.5, 0.5); % up-left from v2
    \vertex [label=right:{\footnotesize 4}] (e2) at (1.1, 0.5); % up-right from v2
    \vertex [label=right:{\footnotesize 5}] (f1) at (2.1, 0.4);
    \vertex [label=right:{\footnotesize 6}] (f2) at (2.1, -0.4);
    \diagram* {
      (i1) -- (v1), (i2) -- (v1),
      (v1) -- [scalar] (v2),
      (v2) -- (e1), (v2) -- (e2),
      (v2) -- [scalar] (v3),
      (v3) -- (f1), (v3) -- (f2),
    };
  \end{feynman}
\end{tikzpicture}
=i\frac{\Lambda^2}{216\Mpl^6}p_{34}^2 ~. 
\end{align}
In these diagrams, the numbers attached to each external line denote the corresponding momentum $ p_i $ with $ i = 1, \ldots, 6 $. 
The propagators for $ \delta \Phi_{\rm E} $ and $ \delta \phi $ are represented by the dashed line and the solid line, respectively. 
Their expressions are derived from the free part of the Lagrangian $ \mathcal{L}_{{\rm E},2} $ in Eq.~\eqref{eq-EF-action-perturbation} and are shown in Eqs.~\eqref{eq-EF-propagator-Phi} and \eqref{eq-EF-propagator-phi}. 
One can also define a canonicalized field $ \delta \phi_{\rm E} \equiv \Lambda/(\sqrt{6}\Mpl)\, \delta \phi $ which coincides with $ \delta \phi_{\rm J} $ due to $ v=0 $. 
The Feynman rules for the $ n $-point vertices where $ n \geq 3 $ can be derived from the interaction terms in $ \mathcal{L}_{{\rm E},3} $, $ \mathcal{L}_{{\rm E},4} $, $ \mathcal{L}_{{\rm E},5} $, and $ \mathcal{L}_{{\rm E},6} $, as shown in Eqs.~\eqref{eq-EF-3pt} to \eqref{eq-EF-6pt}. 
Except for the first diagram in Eq.~\eqref{eq-EF-six-point} which is a six-point contact interaction absent in the Jordan frame, the other five types of diagrams have the same structures as the ones in the Jordan frame, although their contributions are differently calculated with the Einstein frame Lagrangian~\eqref{eq-EF-action-perturbation}.

Summing over all momentum configurations as given in Eqs.~\eqref{eq-EF-sum1} to \eqref{eq-EF-sum5} in Appendix~\ref{app-EF}, the final result for six-point on-shell amplitudes for $ \delta \phi $ in the Einstein frame is 
\begin{equation}
\begin{split}\label{eq-EF-6-point-M}
    \overline{\mathcal{M}}_{\phi\phi\phi\phi\phi\phi}&=\frac{\lambda^2\Lambda^6}{6\Mpl^6}\sum^{10}\frac{1}{p_{ijk}^2}+\frac{\lambda\Lambda^4}{\Mpl^6}\qty(10\xi+\frac{5}{3}-\frac{10}{3}\xi(1-6\xi)-\frac{10}{9}) ~, \\
    &= \frac{\lambda^2\Lambda^6}{6\Mpl^6}\sum^{10}\frac{1}{p_{ijk}^2}+\frac{5\lambda\Lambda^4}{9\Mpl^6}(1+6\xi)^2 ~,
\end{split}
\end{equation}
which is exactly the same as Eq.~\eqref{eq-JF-6-point-M}. 
Our calculation shows explicit consistency of the unitarity violation scales computed from both frames, especially the correct dependence on $ \lambda $ and $ \xi $ in the Jordan frame. 
Our results clarify the importance of the potential in the determination of the cutoff scale in this model and we have properly taken it into account.

\section{Conclusion and outlook}\label{sec-conclusion}

In this paper, we have revisited the problem of determination of the unitarity violation scale in a single scalar field model with a non-minimal coupling with gravity and a quartic potential. 
We have properly taken into account the effects from both the non-minimal coupling and the potential, explicitly showing the consistency of calculation in the Jordan frame and the Einstein frame. 
We can clearly see that our results become trivial when $ \lambda \to 0 $ or $ \xi \to -1/6 $, as expected by physical consideration. 

In previous studies, especially in the Jordan frame, the calculation focuses on the non-minimal coupling term, which leads to a cutoff scale independent of the self-coupling $ \lambda $ of the scalar field. 
This makes the potential look seemingly unimportant in the determination of the cutoff scale. 
However, in the Einstein frame calculation, $ \lambda $ inevitably appears in the results. 
Physics should not depend on the choice of frame, so the two results must coincide if we have properly considered all the ingredients in both frames. 
Our results, therefore, fill this gap and provide a self-consistent understanding of this problem. 

To avoid such ambiguity in the calculation, a manifestly frame-independent method will be beneficial. 
One way is to geometrize the amplitude calculation with the conformal mode, such that the results can be expressed in terms of geometric quantities, while the frame transformation becomes just a coordinate transformation that does not affect the underlying geometry. 
We leave the geometrization as future work.

%%%%%%%%%%%%%%%%%%%%%%%%%%%%%%%%%%%%%%%%%%%%%%%%%%
\section*{Acknowledgements}
%%%%%%%%%%%%%%%%%%%%%%%%%%%%%%%%%%%%%%%%%%%%%%%%%%
We thank Siyao Li for collaboration at the early stage of the project. 
M.\,He was supported by IBS under the project code, IBS-R018-D1. 
M.\,Hong was supported by Grant-in-Aid for JSPS Fellows 23KJ0697. 
K.\,M.\, was supported by JSPS KAKENHI Grant No.\ JP22K14044. 
All the Feynman diagrams in this paper have been produced by using \texttt{TikZ-FeynHand}~\cite{Dohse:2018vqo,Ellis:2016jkw}.

\appendix

\section{Jordan frame}\label{app-JF}

\subsection{Feynman rules}

In the Jordan frame, we can write down the Feynman rules as follows. 
\begin{enumerate}
    \item Propagators: 
\begin{align}
% PhiPhi propagator
    \begin{tikzpicture}[baseline={([yshift=-.5ex]current bounding box.center)}]
  \begin{feynman}
    \vertex (a);
    \vertex [right=1.5cm of a] (b);
    \diagram* {
      (a) -- [scalar] (b),
    };
  \end{feynman}
\end{tikzpicture}
&\equiv \overline{\Delta}^{\Phi\Phi}(p)=\frac{i}{p^2} \label{eq-JF-propagator-Phi}\\
% phiphi propagator
\begin{tikzpicture}[baseline={([yshift=-.5ex]current bounding box.center)}]
  \begin{feynman}
    \vertex (a);
    \vertex [right=1.5cm of a] (b);
    \diagram* {
      (a) -- [plain] (b),
    };
  \end{feynman}
\end{tikzpicture}
&\equiv \overline{\Delta}^{\phi\phi}(p)=-\frac{6\Mpl^2}{\Lambda^2}\frac{i}{p^2} \label{eq-JF-propagator-phi}
\end{align}
    \item Vertices: 
\begin{align}
% 3pt
\begin{tikzpicture}[baseline={([yshift=-.5ex]current bounding box.center)}]
  \begin{feynman}
    \vertex (v);
    \vertex [above left=0.8cm of v] (i1) {\footnotesize $1$};
    \vertex [below left=0.8cm of v] (i2) {\footnotesize $2$};
    \vertex [right=0.8cm of v] (f1) {\footnotesize $3$};
    \diagram* {
      (i1) -- [scalar] (v),
      (i2) -- [plain] (v),
      (f1) -- [plain] (v),
    };
  \end{feynman}
\end{tikzpicture}
&= \frac{i\Lambda}{6\Mpl^2} \left[(1+6\xi)p_1^2-p_2^2-p_3^2\right] \label{eq-JF-3pt}
\end{align}

\begin{align}
% 4pt phiphiphiphi
\begin{tikzpicture}[baseline={([yshift=-.5ex]current bounding box.center)}]
  \begin{feynman}
    \vertex (v);
    \vertex [above left=0.8cm of v] (i1);
    \vertex [below left=0.8cm of v] (i2);
    \vertex [above right=0.8cm of v] (f1);
    \vertex [below right=0.8cm of v] (f2);
    \diagram* {
      (i1) -- (v),
      (i2) -- (v),
      (f1) -- (v),
      (f2) -- (v),
    };
  \end{feynman}
\end{tikzpicture}
&= - \frac{i \lambda \Lambda^4}{6\Mpl^4} \label{eq-JF-4pt-phi4}
\end{align}

\begin{align}
% 4pt phiphiPhiPhi
\begin{tikzpicture}[baseline={([yshift=-.5ex]current bounding box.center)}]
  \begin{feynman}
    \vertex (v);
    \vertex [above left=0.8cm of v] (i1) {\footnotesize $1$};
    \vertex [above right=0.8cm of v] (i2) {\footnotesize $2$};
    \vertex [below left=0.8cm of v] (i3) {\footnotesize $3$};
    \vertex [below right=0.8cm of v] (i4) {\footnotesize $4$};
    \diagram* {
      (i1) -- [plain] (v),
      (i2) -- [plain] (v),
      (i3) -- [scalar] (v),
      (i4) -- [scalar] (v),
    };
  \end{feynman}
\end{tikzpicture}
&= \frac{i}{6\Mpl^2}\left(-p_1^2 - p_2^2 + p_{12}^2 +6\xi(p_3^2+p_4^2)\right) \label{eq-JF-4pt-phiPhi}
\end{align}

\begin{align}
% 5pt
\begin{tikzpicture}[baseline={([yshift=-.5ex]current bounding box.center)}]
  \begin{feynman}
    \vertex (v);
    \vertex (l1) at (162:0.8);
    \vertex (l2) at (90:0.8);
    \vertex (l3) at (18:0.8);
    \vertex (l4) at (306:0.8);
    \vertex (l5) at (234:0.8);
    \diagram* {
      (l1) -- [scalar] (v),
      (l2) -- (v),
      (l3) -- (v),
      (l4) -- (v),
      (l5) -- (v),
    };
  \end{feynman}
\end{tikzpicture}
&= - i \frac{2\lambda\Lambda^3}{3\Mpl^4} \label{eq-JF-5pt}
\end{align}
\end{enumerate}
Here, we have implicitly imposed the momentum conservation $ \sum_{i=1}^N p_i =0 $ where $ N $ is the number of external legs in the corresponding diagrams. 

\subsection{Summation of diagrams}

In Sec.~\ref{sec-JF}, the calculation involves summation of different configurations of momenta for each type of diagram. 
The results are summarized as follows:
\begin{align}
    \sum^{15}\begin{tikzpicture}[baseline={([yshift=-.5ex]current bounding box.center)}]
  \begin{feynman}
    \vertex (v1) at (0,0);
    \vertex (v2) at (0.8,0);
    \vertex (i1) at (-0.5, 0.4);
    \vertex (i2) at (-0.5, -0.4);
    \vertex (f1) at (1.3, 0.6);
    \vertex (f2) at (1.3, 0.2);
    \vertex (f3) at (1.3, -0.2);
    \vertex (f4) at (1.3, -0.6);
    \diagram* {
      (i1) -- (v1), (i2) -- (v1),
      (v1) -- [scalar] (v2),
      (v2) -- (f1), (v2) -- (f2), (v2) -- (f3), (v2) -- (f4),
    };
  \end{feynman}
\end{tikzpicture}
=i\frac{5\lambda\Lambda^4}{3\Mpl^6}(1+6\xi) ~, \label{eq-JF-sum1}
\end{align}
\begin{align}
    \sum^{10}\begin{tikzpicture}[baseline={([yshift=-.5ex]current bounding box.center)}]
  \begin{feynman}
    \vertex (v1) at (0,0);
    \vertex (v2) at (0.8,0);
    \vertex (i1) at (-0.5, 0.4);
    \vertex (i2) at (-0.6, 0);
    \vertex (i3) at (-0.5, -0.4);
    \vertex (f1) at (1.3, 0.4);
    \vertex (f2) at (1.4, 0);
    \vertex (f3) at (1.3, -0.4);
    \diagram* {
      (i1) -- (v1), (i2) -- (v1), (i3) -- (v1),
      (v1) -- (v2),
      (v2) -- (f1), (v2) -- (f2), (v2) -- (f3),
    };
  \end{feynman}
\end{tikzpicture}
=i\frac{\lambda^2\Lambda^6}{6\Mpl^6}\sum^{10}\frac{1}{p_{ijk}^2} ~, \label{eq-JF-sum2}
\end{align}
\begin{align}
\begin{split}
\sum^{60}\begin{tikzpicture}[baseline={([yshift=-.5ex]current bounding box.center)}]
  \begin{feynman}
    \vertex (v1) at (0,0);
    \vertex (v2) at (0.8,0);
    \vertex (v3) at (1.6,0);
    \vertex (i1) at (-0.5, 0.4);
    \vertex (i2) at (-0.5, -0.4);
    \vertex (e1) at (1.2, 0.5); % up-left from v2
    \vertex (f1) at (2.1, 0.4);
    \vertex (f2) at (2.2, 0);
    \vertex (f3) at (2.1, -0.4);
    \diagram* {
      (i1) -- (v1), (i2) -- (v1),
      (v1) -- [scalar] (v2),
      (v2) -- (e1),
      (v2) -- (v3),
      (v3) -- (f1), (v3) -- (f2), (v3) -- (f3),
    };
  \end{feynman}
\end{tikzpicture}
&=-i\frac{5\lambda\Lambda^4}{3\Mpl^6}(1+6\xi)+i\frac{\lambda\Lambda^4}{36\Mpl^6}(1+6\xi)^2\sum\frac{p_{12}^2}{p_{123}^2} ~,\\
&=-i\frac{5\lambda\Lambda^4}{3\Mpl^6}(1+6\xi)+i\frac{5\lambda\Lambda^4}{9\Mpl^6}(1+6\xi)^2 ~, \label{eq-JF-sum3}
\end{split}
\end{align}
\begin{align}
\sum^{90}\begin{tikzpicture}[baseline={([yshift=-.5ex]current bounding box.center)}]
  \begin{feynman}
    \vertex (v1) at (0,0);
    \vertex (v2) at (0.8,0);
    \vertex (v3) at (1.2,0);
    \vertex (v4) at (2.0,0);
    \vertex (i1) at (-0.5, 0.4);
    \vertex (i2) at (-0.5, -0.4);
    \vertex (e1) at (0.4, 0.5); % up-left from v2
    \vertex (e2) at (1.6, 0.5); % up-right from v3
    \vertex (f1) at (2.5, 0.4);
    \vertex (f2) at (2.5, -0.4);
    \diagram* {
      (i1) -- (v1), (i2) -- (v1),
      (v1) -- [scalar] (v2),
      (v2) -- (e1),
      (v2) -- (v3), % plain (solid)
      (v3) -- (e2),
      (v3) -- [scalar] (v4),
      (v4) -- (f1), (v4) -- (f2),
    };
  \end{feynman}
\end{tikzpicture}
=0 ~, \label{eq-JF-sum4}
\end{align}
\begin{align}
\sum^{45}\begin{tikzpicture}[baseline={([yshift=-.5ex]current bounding box.center)}]
  \begin{feynman}
    \vertex (v1) at (0,0);
    \vertex (v2) at (0.8,0);
    \vertex (v3) at (1.6,0);
    \vertex (i1) at (-0.5, 0.4);
    \vertex (i2) at (-0.5, -0.4);
    \vertex (e1) at (0.5, 0.5); % up-left from v2
    \vertex (e2) at (1.1, 0.5); % up-right from v2
    \vertex (f1) at (2.1, 0.4);
    \vertex (f2) at (2.1, -0.4);
    \diagram* {
      (i1) -- (v1), (i2) -- (v1),
      (v1) -- [scalar] (v2),
      (v2) -- (e1), (v2) -- (e2),
      (v2) -- [scalar] (v3),
      (v3) -- (f1), (v3) -- (f2),
    };
  \end{feynman}
\end{tikzpicture}
=0 ~, \label{eq-JF-sum5}
\end{align}
where $ \sum $ represents the summation over all different momentum configurations. 
Here, we have imposed the on-shell conditions $ p_i^2 =0 $ for $ i = 1, 2, \ldots, 6 $, which allows us to derive 
\begin{align}
    \sum_{i,j=1}^n p_i \cdot p_j = \frac{1}{2} \left( \sum_{i=1}^n p_i \right)^2 ~. 
\end{align}
If $ n = 5 $, conservation of total momentum tells us that $ \sum_{i=1}^5 p_i =-p_6 $ such that the square of it vanishes. 
If $ n=6 $, $ \sum_{i=1}^6 p_i =0 $. 
This is useful to simplify the summation of different momentum configurations.

\section{Einstein frame}\label{app-EF}

\subsection{Feynman rules}

The Feynman rules in the Einstein frame are given in the following. 
\begin{enumerate}
    \item Propagators:
\begin{align}
% PhiPhi propagator
    \begin{tikzpicture}[baseline={([yshift=-.5ex]current bounding box.center)}]
  \begin{feynman}
    \vertex (a);
    \vertex [right=1.5cm of a] (b);
    \diagram* {
      (a) -- [scalar] (b),
    };
  \end{feynman}
\end{tikzpicture}
&\equiv \overline{\Delta}^{\Phi\Phi}(p)=\frac{i}{p^2} \label{eq-EF-propagator-Phi}\\
% phiphi propagator
\begin{tikzpicture}[baseline={([yshift=-.5ex]current bounding box.center)}]
  \begin{feynman}
    \vertex (a);
    \vertex [right=1.5cm of a] (b);
    \diagram* {
      (a) -- [plain] (b),
    };
  \end{feynman}
\end{tikzpicture}
&\equiv \overline{\Delta}^{\phi\phi}(p)=-\frac{6\Mpl^2}{\Lambda^2}\frac{i}{p^2} \label{eq-EF-propagator-phi} 
\end{align}
    \item Vertices: 
\begin{align}
% 3pt
\begin{tikzpicture}[baseline={([yshift=-.5ex]current bounding box.center)}]
  \begin{feynman}
    \vertex (v);
    \vertex [above left=0.8cm of v] (i1) {\footnotesize $1$};
    \vertex [below left=0.8cm of v] (i2) {\footnotesize $2$};
    \vertex [right=0.8cm of v] (f1) {\footnotesize $3$};
    \diagram* {
      (i1) -- [scalar] (v),
      (i2) -- [plain] (v),
      (f1) -- [plain] (v),
    };
  \end{feynman}
\end{tikzpicture}
&=\frac{i\Lambda}{6\Mpl^2}(p_1^2-p_2^2-p_3^2) \label{eq-EF-3pt}
\end{align}

\begin{align}
% 4pt phiphiphiphi
\begin{tikzpicture}[baseline={([yshift=-.5ex]current bounding box.center)}]
  \begin{feynman}
    \vertex (v);
    \vertex [above left=0.8cm of v] (i1) {\footnotesize $1$};
    \vertex [below left=0.8cm of v] (i2) {\footnotesize $2$};
    \vertex [above right=0.8cm of v] (f1) {\footnotesize $3$};
    \vertex [below right=0.8cm of v] (f2) {\footnotesize $4$};
    \diagram* {
      (i1) -- [plain](v),
      (i2) -- [plain](v),
      (f1) -- [plain](v),
      (f2) -- [plain](v),
    };
  \end{feynman}
\end{tikzpicture}
&=\frac{i\xi\Lambda^2}{6\Mpl^4}(1-6\xi)(p_1^2+p_2^2+p_3^2+p_4^2)-\frac{i\lambda\Lambda^4}{6\Mpl^4} \label{eq-EF-4pt-phi4}
\end{align}

\begin{align}
% 4pt phiphiPhiPhi
\begin{tikzpicture}[baseline={([yshift=-.5ex]current bounding box.center)}]
  \begin{feynman}
    \vertex (v);
    \vertex [above left=0.8cm of v] (i1) {\footnotesize $1$};
    \vertex [above right=0.8cm of v] (i2) {\footnotesize $2$};
    \vertex [below left=0.8cm of v] (i3) {\footnotesize $3$};
    \vertex [below right=0.8cm of v] (i4) {\footnotesize $4$};
    \diagram* {
      (i1) -- [plain] (v),
      (i2) -- [plain] (v),
      (i3) -- [scalar] (v),
      (i4) -- [scalar] (v),
    };
  \end{feynman}
\end{tikzpicture}
&=-\frac{i}{6\Mpl^2}\left(p_1^2+p_2^2-p_{12}^2\right) \label{eq-EF-4pt-phiPhi}
\end{align}

\begin{align}
% 5pt
\begin{tikzpicture}[baseline={([yshift=-.5ex]current bounding box.center)}]
  \begin{feynman}
    \vertex (v);
    \vertex (l1) at (162:0.8) {\footnotesize $1$};
    \vertex (l2) at (90:0.8) {\footnotesize $2$};
    \vertex (l3) at (18:0.8) {\footnotesize $3$};
    \vertex (l4) at (306:0.8) {\footnotesize $4$};
    \vertex (l5) at (234:0.8) {\footnotesize $5$};
    \diagram* {
      (l1) -- [scalar] (v),
      (l2) -- [plain](v),
      (l3) -- [plain](v),
      (l4) -- [plain](v),
      (l5) -- [plain](v),
    };
  \end{feynman}
\end{tikzpicture}
&=\frac{i\xi\Lambda}{3\Mpl^4}(1-6\xi)(-p_1^2+p_2^2+p_3^2+p_4^2+p_5^2)-i\frac{2\lambda\Lambda^3}{3\Mpl^4} \label{eq-EF-5pt}
\end{align}

\begin{align}
%6pt
\begin{tikzpicture}[baseline={([yshift=-.5ex]current bounding box.center)}]
  \begin{feynman}
    \vertex (v);
    \vertex (l1) at (150:0.8) {\footnotesize $1$};
    \vertex (l2) at (90:0.8) {\footnotesize $2$};
    \vertex (l3) at (30:0.8) {\footnotesize $3$};
    \vertex (l4) at (330:0.8) {\footnotesize $4$};
    \vertex (l5) at (270:0.8) {\footnotesize $5$};
    \vertex (l6) at (210:0.8) {\footnotesize $6$};
    \diagram* {
      (l1) -- [plain](v),
      (l2) -- [plain](v),
      (l3) -- [plain](v),
      (l4) -- [plain](v),
      (l5) -- [plain](v),
      (l6) -- [plain](v),
    };
  \end{feynman}
\end{tikzpicture}
&=-i\frac{2\xi^2\Lambda^2}{\Mpl^6}(1-12\xi)(p_1^2+p_2^2+p_3^2+p_4^2+p_5^2+p_6^2)+i\frac{10\xi\lambda\Lambda^4}{\Mpl^6} \label{eq-EF-6pt}
\end{align}
\end{enumerate}

\subsection{Summation of diagrams}

In Sec.~\ref{sec-EF}, the calculation involves summation of different configurations of momenta for each type of diagram. 
The results are summarized as follows: 
\begin{align}
    \sum^{15}\begin{tikzpicture}[baseline={([yshift=-.5ex]current bounding box.center)}]
  \begin{feynman}
    \vertex (v1) at (0,0);
    \vertex (v2) at (0.8,0);
    \vertex (i1) at (-0.5, 0.4);
    \vertex (i2) at (-0.5, -0.4);
    \vertex (f1) at (1.3, 0.6);
    \vertex (f2) at (1.3, 0.2);
    \vertex (f3) at (1.3, -0.2);
    \vertex (f4) at (1.3, -0.6);
    \diagram* {
      (i1) -- (v1), (i2) -- (v1),
      (v1) -- [scalar] (v2),
      (v2) -- (f1), (v2) -- (f2), (v2) -- (f3), (v2) -- (f4),
    };
  \end{feynman}
\end{tikzpicture}
=i\frac{5\lambda\Lambda^4}{3\Mpl^6} ~, \label{eq-EF-sum1}
\end{align}

\begin{align}
    \sum^{10}\begin{tikzpicture}[baseline={([yshift=-.5ex]current bounding box.center)}]
  \begin{feynman}
    \vertex (v1) at (0,0);
    \vertex (v2) at (0.8,0);
    \vertex (i1) at (-0.5, 0.4);
    \vertex (i2) at (-0.6, 0);
    \vertex (i3) at (-0.5, -0.4);
    \vertex (f1) at (1.3, 0.4);
    \vertex (f2) at (1.4, 0);
    \vertex (f3) at (1.3, -0.4);
    \diagram* {
      (i1) -- (v1), (i2) -- (v1), (i3) -- (v1),
      (v1) -- (v2),
      (v2) -- (f1), (v2) -- (f2), (v2) -- (f3),
    };
  \end{feynman}
\end{tikzpicture}
=-i\frac{10\xi\lambda\Lambda^4}{3\Mpl^6}(1-6\xi)+i\frac{\lambda^2\Lambda^6}{6\Mpl^6}\sum^{10}\frac{1}{p_{ijk}^2} ~, \label{eq-EF-sum2}
\end{align}

\begin{align}
\sum^{60}\begin{tikzpicture}[baseline={([yshift=-.5ex]current bounding box.center)}]
  \begin{feynman}
    \vertex (v1) at (0,0);
    \vertex (v2) at (0.8,0);
    \vertex (v3) at (1.6,0);
    \vertex (i1) at (-0.5, 0.4);
    \vertex (i2) at (-0.5, -0.4);
    \vertex (e1) at (1.2, 0.5); % up-left from v2
    \vertex (f1) at (2.1, 0.4);
    \vertex (f2) at (2.2, 0);
    \vertex (f3) at (2.1, -0.4);
    \diagram* {
      (i1) -- (v1), (i2) -- (v1),
      (v1) -- [scalar] (v2),
      (v2) -- (e1),
      (v2) -- (v3),
      (v3) -- (f1), (v3) -- (f2), (v3) -- (f3),
    };
  \end{feynman}
\end{tikzpicture}
=-i\frac{10\lambda\Lambda^4}{9\Mpl^6} ~, \label{eq-EF-sum3}
\end{align}

\begin{align}
\sum^{90}\begin{tikzpicture}[baseline={([yshift=-.5ex]current bounding box.center)}]
  \begin{feynman}
    \vertex (v1) at (0,0);
    \vertex (v2) at (0.8,0);
    \vertex (v3) at (1.2,0);
    \vertex (v4) at (2.0,0);
    \vertex (i1) at (-0.5, 0.4);
    \vertex (i2) at (-0.5, -0.4);
    \vertex (e1) at (0.4, 0.5); % up-left from v2
    \vertex (e2) at (1.6, 0.5); % up-right from v3
    \vertex (f1) at (2.5, 0.4);
    \vertex (f2) at (2.5, -0.4);
    \diagram* {
      (i1) -- (v1), (i2) -- (v1),
      (v1) -- [scalar] (v2),
      (v2) -- (e1),
      (v2) -- (v3), % plain (solid)
      (v3) -- (e2),
      (v3) -- [scalar] (v4),
      (v4) -- (f1), (v4) -- (f2),
    };
  \end{feynman}
\end{tikzpicture}
=0 ~,\label{eq-EF-sum4}
\end{align}

\begin{align}
\sum^{45}\begin{tikzpicture}[baseline={([yshift=-.5ex]current bounding box.center)}]
  \begin{feynman}
    \vertex (v1) at (0,0);
    \vertex (v2) at (0.8,0);
    \vertex (v3) at (1.6,0);
    \vertex (i1) at (-0.5, 0.4);
    \vertex (i2) at (-0.5, -0.4);
    \vertex (e1) at (0.5, 0.5); % up-left from v2
    \vertex (e2) at (1.1, 0.5); % up-right from v2
    \vertex (f1) at (2.1, 0.4);
    \vertex (f2) at (2.1, -0.4);
    \diagram* {
      (i1) -- (v1), (i2) -- (v1),
      (v1) -- [scalar] (v2),
      (v2) -- (e1), (v2) -- (e2),
      (v2) -- [scalar] (v3),
      (v3) -- (f1), (v3) -- (f2),
    };
  \end{feynman}
\end{tikzpicture}
=0  ~. \label{eq-EF-sum5}
\end{align}

%%%%%%%%%%%%%%%%%%%%%%
\bibliographystyle{utphys}
\bibliography{ref}
%%%%%%%%%%%%%%%%%%%%%%

\end{document}